\documentclass[10pt,twocolumn]{article}
\usepackage[utf8]{inputenc}
\usepackage{mathptmx} 
\usepackage[T1]{fontenc} 
\usepackage{amsmath,amssymb}
\usepackage[margin=1in]{geometry}
\usepackage{hyperref}
\usepackage{multicol}
\usepackage{etoolbox}
\usepackage{caption}
\usepackage{svg}
\usepackage{float}
\usepackage{natbib}
\usepackage{soul,color}
\usepackage{titlesec}
\titlespacing*{\section}{0pt}{3.5ex}{1.5ex} 

\makeatletter
\renewcommand{\maketitle}{%
  \begin{center}%
    {\fontsize{14}{17}\selectfont\textbf{\@title}}%
  \end{center}%
}
\makeatother

\usepackage{titlesec}
\titleformat*{\section}{\fontsize{11}{13}\selectfont\bfseries}
\titleformat*{\subsection}{\fontsize{11}{13}\selectfont\bfseries}
\titleformat*{\subsubsection}{\fontsize{11}{13}\selectfont\bfseries}

\usepackage{natbib}

\setcitestyle{numbers,square}

\begin{document}

\pagestyle{plain}

\twocolumn[{
\begin{center}
\textit{7\textsuperscript{th} International Conference on Advances in Signal Processing and Artificial Intelligence \\(ASPAI' 2025), 8-10 April 2025, Innsbruck, Austria}
\end{center}

{\fontsize{14}{17}\selectfont
\begin{center}
\textbf{Deep Jansen-Rit Parameter Inference for Model-Driven Analysis of Brain Activity}
\end{center}
}
\vspace{-0.6cm}
\begin{center}
\large{D. Tilwani\textsuperscript{1-4} and C. O'Reilly\textsuperscript{1-4}}

\vspace{0.1cm}
\normalsize{\textsuperscript{1} Department of Computer Science and Engineering, University of South Carolina, Columbia, SC, USA}

\vspace{0.1cm}
\normalsize{\textsuperscript{2} Artificial Intelligence Institute, University of South Carolina, Columbia, SC, USA}

\vspace{0.1cm}
\normalsize{\textsuperscript{3} Carolina Autism and Neurodevelopment Research Center, University of South Carolina, Columbia, SC, USA}

\vspace{0.1cm}
\normalsize{\textsuperscript{4} Institute for Mind and Brain, University of South Carolina, Columbia, SC, USA}

\normalsize{E-mail: \underline{dtilwani@mailbox.sc.edu}, christian.oreilly@sc.edu}
\end{center}

\vspace{-0.1cm}
\noindent\rule{\textwidth}{0.2pt}

\vspace{0.3cm}

\textbf{Summary:} Accurately modeling effective connectivity (EC) is critical for understanding how the brain processes and integrates sensory information. Yet, it remains a formidable challenge due to complex neural dynamics and noisy measurements such as those obtained from the electroencephalogram (EEG). Model-driven EC infers local (within a brain region) and global (between brain regions) EC parameters by fitting a generative model of neural activity onto experimental data. This approach offers a promising route for various applications, including investigating neurodevelopmental disorders. However, current approaches fail to scale to whole-brain analyses and are highly noise-sensitive. In this work, we employ three deep-learning architectures—a transformer, a long short-term memory (LSTM) network, and a convolutional neural network and bidirectional LSTM (CNN-BiLSTM) network—for inverse modeling and compare their performance with simulation-based inference in estimating the Jansen-Rit neural mass model (JR-NMM) parameters from simulated EEG data under various noise conditions. We demonstrate a reliable estimation of key local parameters, such as synaptic gains and time constants. However, other parameters like local JR-NMM connectivity cannot be evaluated reliably from evoked-related potentials (ERP). We also conduct a sensitivity analysis to characterize the influence of JR-NMM parameters on ERP and evaluate their learnability. Our results show the feasibility of deep-learning approaches to estimate the subset of learnable JR-NMM parameters.  

\vspace{0.2cm}
\textbf{Keywords:} EEG, Neural Mass Model, LSTM, Transformers, Simulation-based Inference, Model-Driven Analysis

\vspace{-1mm}
\noindent\rule{\textwidth}{0.2pt}
\vspace{0.5mm}
}]

\section{Introduction}
Neural mass models (NMMs) have emerged as powerful computational tools for simulating collective neuronal behavior at a mesoscopic scale, offering a mathematically tractable approach to modeling brain dynamics for the past five decades \cite{
lopes1974model, 
Wilson1972ExcitatoryAI}. These models provide a crucial bridge between microscopic neural activity and macroscopic neuroimaging signals, enabling quantitative analysis of normal brain functions and pathophysiological conditions \cite{deco2008dynamic}. NMMs create computationally efficient abstractions that can be simulated at scale across brain regions by capturing the dynamics of the average activity of neuronal populations as systems of coupled differential equations. Integrating NMMs with computational data analysis pipelines has enabled significant advances in neuroimaging \cite{sotero2007realistically, articleDavidNMM}. Modern implementations of these models allow researchers to simulate the transition between stable and pathological brain states, providing a basis for biomarker development and quantitative analysis of neurological disorders. NMMs can be particularly effective for modeling epileptiform activity \cite{wendling2016computational}, oscillatory disturbances \cite{kuhlmann2016neural}, and state transitions in disorders like Alzheimer's disease \cite{stefanovski2021bridging}.

Effective connectivity (EC) captures the causal influence that neural systems exert on one another. Unlike functional connectivity, which merely describes statistical dependencies, EC captures the directional influence of one neural population on another. Dynamic causal modeling (DCM) \cite{articleDavidNMM} represents a principled algorithmic approach to estimating these causal relationships from neural signals. This approach frames the problem as a Bayesian model inversion and explains observed data through a generative model of coupled neural masses. Bayesian techniques and, hence, DCM \cite{greenberg2019automatic, cranmer2020frontier, gelman2006data}) are computationally demanding and, therefore, are generally inadequate for handling large datasets or real-time analysis.

Simulation-based inference (SBI) has recently emerged as a computationally efficient approach for parameter estimation in complex dynamical systems \cite{tejero-cantero2020sbi}. SBI algorithms, including Sequential Neural Posterior Estimation (SNPE) \cite{papamakarios2016fast}, leverage advances in amortized inference to approximate posterior distributions without requiring explicit likelihood calculations. These methods can handle complex forward models but face diminishing performance as the parameter spaces increase, creating computational bottlenecks for whole-brain analyses.

Deep learning architectures present a promising paradigm for parameter recovery in complex dynamical systems. Recent advances in neural network architectures — particularly sequence models like Transformers \cite{anwar2024transformers} and Long Short-Term Memory (LSTM) \cite{yu2019review} networks — can efficiently learn the mapping between model parameters and observable outputs. These networks compress the high-dimensional functional relationship between parameters and outputs, enabling rapid inference once trained. Recent research has successfully applied these techniques to connectome-based NMMs for functional magnetic resonance imaging (fMRI) \cite{GriffithsDCMfmri}, suggesting broader applicability to neuroimaging data analysis.

In this paper, we systematically evaluate four approaches for parameter inference in the Jansen-Rit Neural Mass Model (JR-NMM): (1) an EEGTransformer architecture using multi-head attention mechanisms and Transformer Encoder components only, (2) an LSTM network specialized for sequential data processing, (3) a convolutional neural network and bidirectional LSTM (CNN-BiLSTM), and (4) the SBI approach using SNPE. We benchmark these algorithms' performance in estimating nine JR-NMM parameters (as defined later, in Table \ref{tab:parms}) from EEG data under varying noise conditions. Through comprehensive computational experiments and sensitivity analyses, we establish the relative strengths of each approach and identify which parameters can be recovered from EEG signals reliably. This computational benchmarking study lays essential groundwork for developing robust algorithms that infer local and global parameters from empirical EEG recordings.

\section{Background}

Brain dynamics span multiple spatial and temporal scales. Accordingly, its computational modeling ranges from detailed biophysical simulations of individual neurons to abstract population-level models of regional activity \cite{articleDavidNMM, 
articleNMF, sadeghi2020dynamic}. NMMs occupy a crucial middle ground in this landscape, providing computationally tractable abstractions of neural population dynamics anchored in neurophysiological mechanisms. These models distill the essential properties of neural circuits while remaining amenable to efficient numerical simulation, making them valuable tools for neuroinformatics and computational neuroscience.

When formalized in 1995 \cite{JansenRit1995}, JR-NMM represented a significant advance in mesoscopic brain modeling. This model extended earlier mathematical formulations by Lopes da Silva \cite{lopes1974model} by implementing a cortical column as a system of coupled differential equations representing interacting populations of pyramidal cells, excitatory interneurons, and inhibitory interneurons. This architecture enables the simulation of various EEG rhythms and event-related potentials (ERPs) through appropriate parameter configurations. The computational efficiency and biological plausibility of models like JR-NMM made them foundational components in whole-brain simulation platforms, including frameworks like The Virtual Brain (TVB) \cite{sanz2013virtual}, neurolib \cite{cakan2021}, and FastDMF \cite{herzog2024neural}.

The JR-NMM is formulated as a system of three second-order ordinary differential equations (ODEs) and generally transformed into six first-order ODEs. This formulation enables efficient numerical integration using standard ODE solvers such as Euler or Runge-Kutta methods. The model parameters — which include synaptic gains ($A_e$, $A_i$), time constants ($b_e$, $b_i$), connectivity strengths ($a_1$-$a_4$), and an overall scaling factor ($C$) — determine the dynamics of the system. Variations in these parameters can produce diverse computational behaviors, including fixed points, limit cycles, and chaotic attractors, corresponding to different patterns observed in the electroencephalogram (EEG) \cite{spiegler2011modeling}.

Challenges in parameter estimation in NMMs stem from several algorithmic and mathematical factors. First, the nonlinear nature of these models creates a complex relationship between parameters and observable outputs. Second, the models exhibit partial identifiability, where multiple parameter combinations can produce nearly identical outputs, making their fitting a mathematically ill-posed inverse problem. Third, the high-dimensional parameter space creates a computationally intensive search problem that scales poorly with traditional optimization methods. Finally, the stochastic nature of neural recordings due to measurement noise and unmeasured inputs introduces additional complexity.

Conventional approaches to NMM parameter estimation have primarily relied on Bayesian inference. DCM \cite{articleDavidNMM} implements a variational Bayesian algorithm to approximate posterior distributions over model parameters. While mathematically elegant, this implementation relies on gradient-based optimization under a linear approximation of model dynamics, limiting its applicability to large-scale models or highly nonlinear systems \cite{sadeghi2020dynamic}. The computational complexity of this algorithm increases exponentially with the number of brain regions, making whole-brain analyses computationally prohibitive without substantial high-performance computing resources.

Recent advances in SBI \cite{tejero-cantero2020sbi} have introduced new paradigms for parameter estimation in complex dynamical systems. Unlike traditional Bayesian methods that require explicit likelihood functions, SBI algorithms like SNPE approximate posterior distributions directly through repeated simulations \cite{papamakarios2016fast}. These methods leverage neural density estimators to learn the conditional distribution of parameters given observed data. While SBI provides greater flexibility for complex forward models, the rigid requirements for the experimental data to closely match the generated simulation data in all aspects (i.e., noise characteristics, free parameters, exact priors, and other underlying properties) limits this approach \cite{wang2023sbi++}.

Deep learning architectures offer alternative strategies for parameter recovery in NMMs. Transformer models built on self-attention mechanisms can capture long-range dependencies in time series without the sequential constraints of recurrent networks. Meanwhile, LSTM provides specialized computational structures for processing sequential information through gated memory cells that can retain information over an extended period. Both architectures can effectively learn complex mappings between model parameters and observable outputs when trained on sufficiently large simulation datasets.

The study of effective connectivity (EC) extends beyond methodological considerations, addressing fundamental questions about information processing in neural systems. EC quantifies the causal influence neural populations exert on each other, providing insights into functional integration and segregation within the brain. Through analyses of neuroimaging data, disruptions in EC have been implicated in various neurological and psychiatric conditions. Developing computationally efficient and robust methods for EC inference could enhance both the theoretical understanding of brain organization and practical applications in clinical neuroscience. Our work extends the computational neuroscience literature by systematically benchmarking four distinct algorithmic approaches to parameter inference in the JR-NMM. We establish each method's relative computational efficiency and accuracy by comparing EEGTransformer, LSTM, and CNN-BiLSTM architectures, and an SNPE-based simulation approach across different noise conditions. This comparative analysis provides crucial insights into which parameters are most reliably recoverable from EEG data. The resulting benchmark offers a foundation for developing more sophisticated algorithms to infer local (within a brain region) and global (between brain regions) parameters in whole-brain simulations based on NMMs.

\section{Methods}

\textbf{Neural Mass Model:} We implemented the JR-NMM \cite{JansenRit1995} to simulate cortical activity. It is governed by the system of coupled differential equations (1). In this system, indices 0, 2, and 4 represent the pyramidal, excitatory, and inhibitory neuron populations, respectively. The output signal $y(t)= a_2y_2 - a_4y_4$ represents the difference between the pyramidal's excitatory and inhibitory postsynaptic potentials. This value is a proxy for EEG sources because EEG is thought to reflect mainly the postsynaptic potentials in the apical dendrites of pyramidal cells \cite{nunez2006electric}. \autoref{tab:parms} shows the default parameter values and ranges used in our simulations based on physiologically plausible values from previous studies \cite{JansenRit1995, articleDavidNMM}.

\vspace{-6mm}
\begin{align}
\dot{y}_0 &= y_1,\ \ \  \dot{y}_2 = y_3, \ \ \ \ \dot{y}_4 = y_5, \nonumber \\
\dot{y}_1 &= A_e b_e S(I_p + a_2 y_2 - a_4 y_4) - 2b_e y_1 - b_e^2 y_0, \nonumber \\
\dot{y}_3 &= A_e b_e S(a_1 y_0)  - 2b_e y_3 - b_e^2 y_2, \\ 
\dot{y}_5 &=A_i b_i S(I_i + a_3 y_0) - 2b_i y_5 - b_i^2 y_4. \nonumber
\end{align}
\vspace{-6mm}

\begin{table}[ht]
\centering
\footnotesize
\caption{Standard parameter settings for Jansen-Rit model. Only the value from parameters with a range is estimated in this study.}
\vspace{-3mm}
\begin{tabular}{lccc}
\hline
\textbf{Parameter} & \textbf{Description} & \textbf{Default} & \textbf{Range} \\
\hline
$A_e$ & Excitatory gain & 3.25 mV & 2.6--9.75 mV \\
$A_i$ & Inhibitory gain & 22 mV & 17.6--110.0 mV \\
$b_e$ & Excit. time const. & 100 s$^{-1}$ & 5--150 s$^{-1}$ \\
$b_i$ & Inhib. time const. & 50 s$^{-1}$ & 25--75 s$^{-1}$ \\
$C$ & Connect. const. & 135 & 65--1350 \\
$a_1$ & Connect. param. & 1.0 & 0.5--1.5 \\
$a_2$ & Connect. param. & 0.8 & 0.4--1.2 \\
$a_3$ & Connect. param. & 0.25 & 0.125--0.375 \\
$a_4$ & Connect. param. & 0.25 & 0.125--0.375 \\
$v_{max}$ & Max firing rate & 5 s$^{-1}$ & -- \\
$v_0$ & Firing threshold & 6 mV & -- \\
$r$ & Sigmoid steepness & 0.56 & -- \\
\hline
\end{tabular}
\label{tab:parms}
\end{table}
\vspace{-2mm}

\textbf{Simulation Protocol:} The simulations were performed with a temporal resolution of 1 ms. Each simulation consisted of 1) an 800 ms transient period to allow the system to reach a steady state, 2) a 200 ms baseline period before stimulus onset, and 3) stimulus events with an inter-stimulus interval of 1 second, presented 60 times. We modeled the stimulus with a 50 ms wide rectangular pulse with a 60 mV ampltiude for pyramidal cells and a 30 mV amplitude for inhibitory interneurons. Moran et al. (2013)\cite{moran2013neural}, we incorporated voltage-dependent synaptic mechanisms in inhibitory interneurons to better capture feedforward inhibition dynamics.  The primary differential equations were solved using a forward Euler integration method. 

\textbf{From Neural Activity to EEG:} We implemented an EEG forward model to transform NMM output into realistic EEG signals. The output signal from the JR-NMM (difference between pyramidal and inhibitory potentials) was scaled by a gain factor of 10$^{-6}$ to account for the scale simulated signals to physiologically realistic EEG amplitudes (10-100$\mu$V). This signal was then used as a source located to a specific cortical region ("caudalmiddlefrontal-lh") using the FreeSurfer average (fsaverage) brain template, as defined in MNE-Python. A forward solution was computed using the Boundary Element Method (BEM) to model volume conduction. 
Sensor projections were created for a standard 64-channel BioSemi EEG montage. To test the robustness of our inference algorithms under different signal-to-noise ratio (SNR) conditions, we added noise to the simulated EEG signals using a scaling approach. We scaled the \textit{ad hoc} noise covariance matrix, generated using MNE's \textit{make\_ad\_hoc\_cov function}, with a factor varying from 0 to 1 in steps of 0.1 to simulate different noise levels added to the synthetic EEG signals. A noise factor of 0 represented noise-free simulations, while a factor of 1.0 applied the full noise covariance, resulting in realistic noise levels comparable to empirical recordings. ERPs were extracted by creating epochs around each stimulus event (from -200 ms to 1000 ms), applying baseline correction using the pre-stimulus interval (-200 to 0 ms), and averaging all 60 trials. 

\vspace{-4mm}
\begin{figure}[ht]
    \centering
    \includegraphics[width=\columnwidth, trim=0 550 0 0,clip]{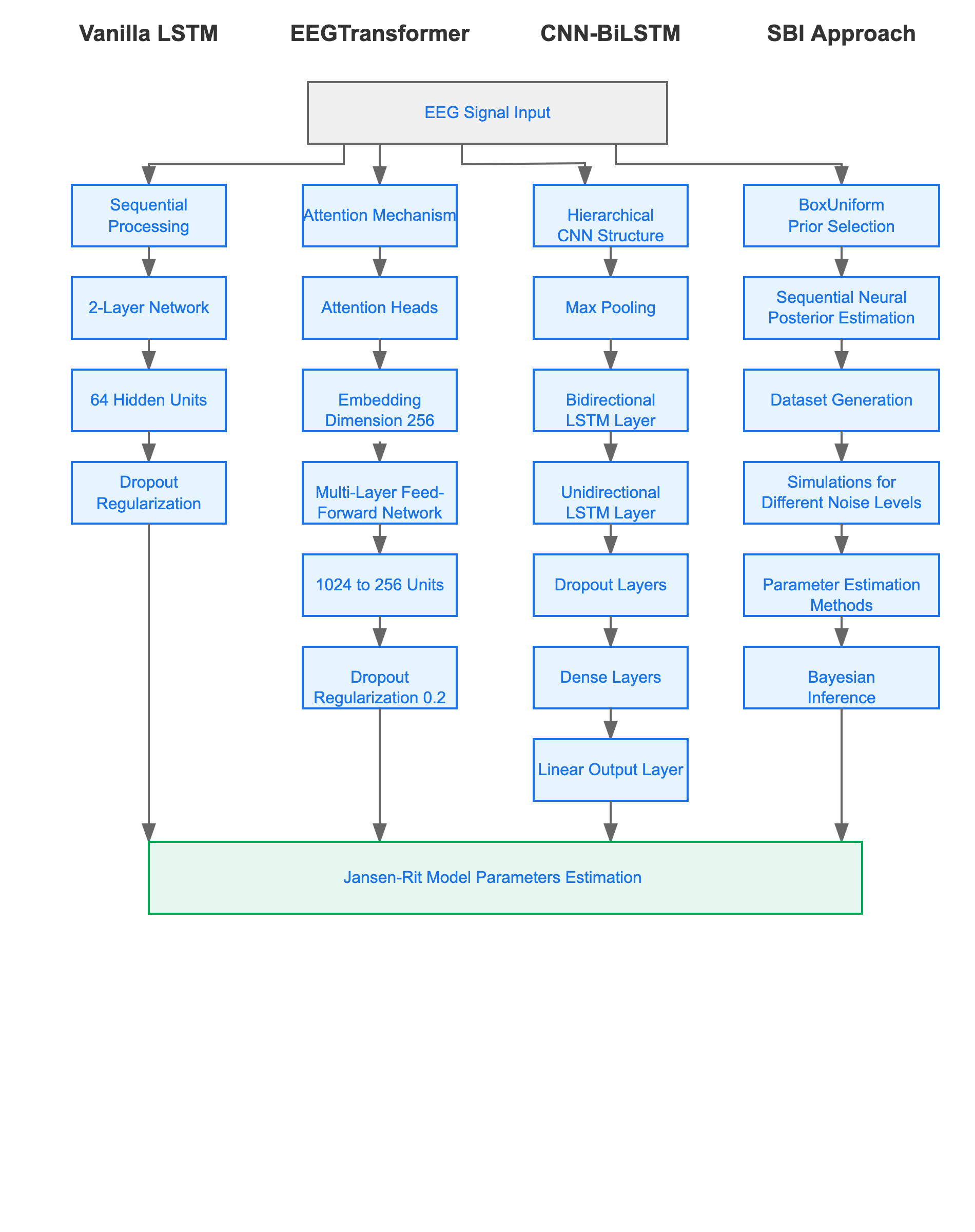}
    \vspace{-7mm}
    \caption{Flow diagram for the inference of $A_e$, $A_i$, $b_e$, $b_i$, $a_1$-$a_4$, and $C$ using the EEGTransformer, LSTM, CNN-BiLSTM, and SBI methods. The JR model generates simulated EEG for training and testing.}
    \label{fig:parameter_inference}
    \vspace{-3mm}
\end{figure}

\textbf{Parameter Inference Models:} We implemented four distinct approaches for inferring the JR-NMM parameters from EEG data, as summarized in \autoref{fig:parameter_inference}. We developed the EEGTransformer model based on Transformer Encoder components only, which processes EEG signals through an attention-based architecture with eight attention heads, an embedding dimension of 256, and a multi-layer feed-forward network (1024→256) with a dropout regularization (0.2). The Vanilla LSTM model treats EEG data as sequential information using a 2-layer network with 64 hidden units and dropout regularization (0.2). The CNN-BiLSTM integrates a CNN for feature extraction with bidirectional LSTMs for temporal dependency analysis. It employs a hierarchical 1D CNN structure (filters: 32→64→128, kernels: 7→5→3) with max pooling, followed by bidirectional and unidirectional LSTM layers (128 units each), multiple dropout layers (0.3), and dense layers (128→64) culminating in a linear output layer. This architecture was designed to capture both spatial patterns across channels and complex temporal dynamics in ERP signals. All neural network models were trained for 50 epochs with batch size 32. The SBI approach uses a BoxUniform prior and sequential neural posterior estimation. To train, we generated a dataset of 1000 simulations per noise level. All four methods were used to estimate the same 9 Jansen-Rit neural mass model parameters ($A_e$, $A_i$, $b_e$, $b_i$, $a_1$-$a_4$, $C$).

\textbf{Data Processing Pipeline:} We adopted a supervised regression approach for parameter inference, following standard deep learning practices \cite{cranmer2020frontier}. \autoref{tab:datapreprocessing} outlines our processing pipeline, designed to optimize neural network training for high-dimensional EEG data.
\begin{table}[ht]
\centering
\footnotesize
\caption{Data processing steps}
\vspace{-3mm}
\begin{tabular}{lp{4.5cm}}
\hline
\textbf{Stage} & \textbf{Details} \\
\hline
Database & 1000 simulations, 64-channel EEG, \\
Creation & 60 trials per simulation \\
\hline
Normalization & JR-NMM parameters normalized to [0,1] \\
\hline
Data Split & 80\% training, 10\% validation, \\
 & 10\% testing (random state=68) \\
\hline
Evaluation & Pearson correlation \\
\hline
\end{tabular}
\label{tab:datapreprocessing}
\end{table}

\textbf{Implementation Details:} Models were implemented in Python 3.12 with PyTorch 1.9. Jansen-Rit neural simulations were mapped to scalp EEG through an EEG forward model implemented using MNE-Python 1.9.0. Additional libraries included xarray for data storage, tqdm for progress tracking, seaborn for visualization, and scikit-learn for preprocessing and evaluation. 

\textbf{Sensitivity Analysis:} We conducted a sensitivity analysis to quantify how variations in Jansen-Rit parameters affect simulated EEG. For each parameter ($A_e$, $A_i$, $b_e$, $b_i$, $a_1$-$a_4$, $C$), we performed 200 simulations while varying the target parameter from the range mentioned in \autoref{tab:parms}, keeping all other parameters fixed.
We generated ERPs for each parameter configuration without noise. This approach allowed us to isolate each parameter's unique influence on EEG output. For quantitative assessment, we quantified parameter sensitivity through three metrics:
\begin{itemize}
    \item Raw evoked potentials: $\mathrm{ERP}(p_i, t)$, where $p_i$ is the parameter, $t$ is time.
    \item Deviations from mean: $\Delta\mathrm{ERP}(p_i, t) = \mathrm{ERP}(p_i, t) - \frac{1}{N}\sum_{j=1}^{N} \mathrm{ERP}(p_j, t)$, with $N=200$ parameter values.
    \item Gradient with respect to parameter: $\nabla_p \mathrm{ERP}(p_i, t) = \left.\frac{\partial \mathrm{ERP}(p, t)}{\partial p}\right|_{p=p_i}$.
\end{itemize}
These metrics were visualized as heatmaps showing their variation depending on parameter values and time. 

\section{Results}
\subsection{Results of Sensitivity Analysis}

\begin{figure}[!ht]
    \centering
    \includegraphics[width=\columnwidth]{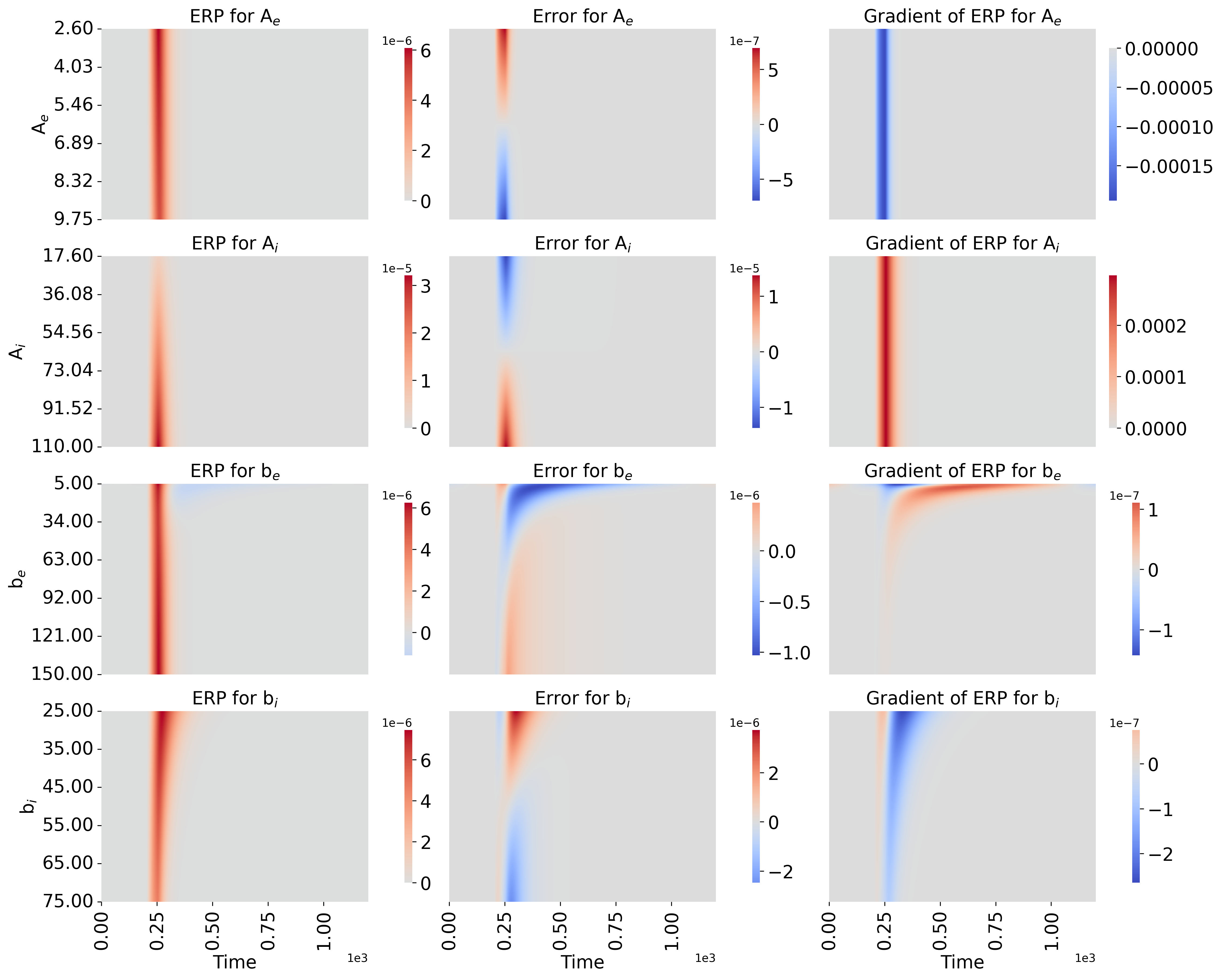}
    \caption{Sensitivity analysis for synaptic parameters ($A_e$, $A_i$, $b_e$, $b_i$) showing the ERP amplitude (left), error (middle), and gradient (right). 
    }
    \label{fig:sensitivity_group1}
    \vspace{-2mm}
\end{figure}

\begin{figure}[!ht]
    \centering
    \includegraphics[width=\columnwidth]{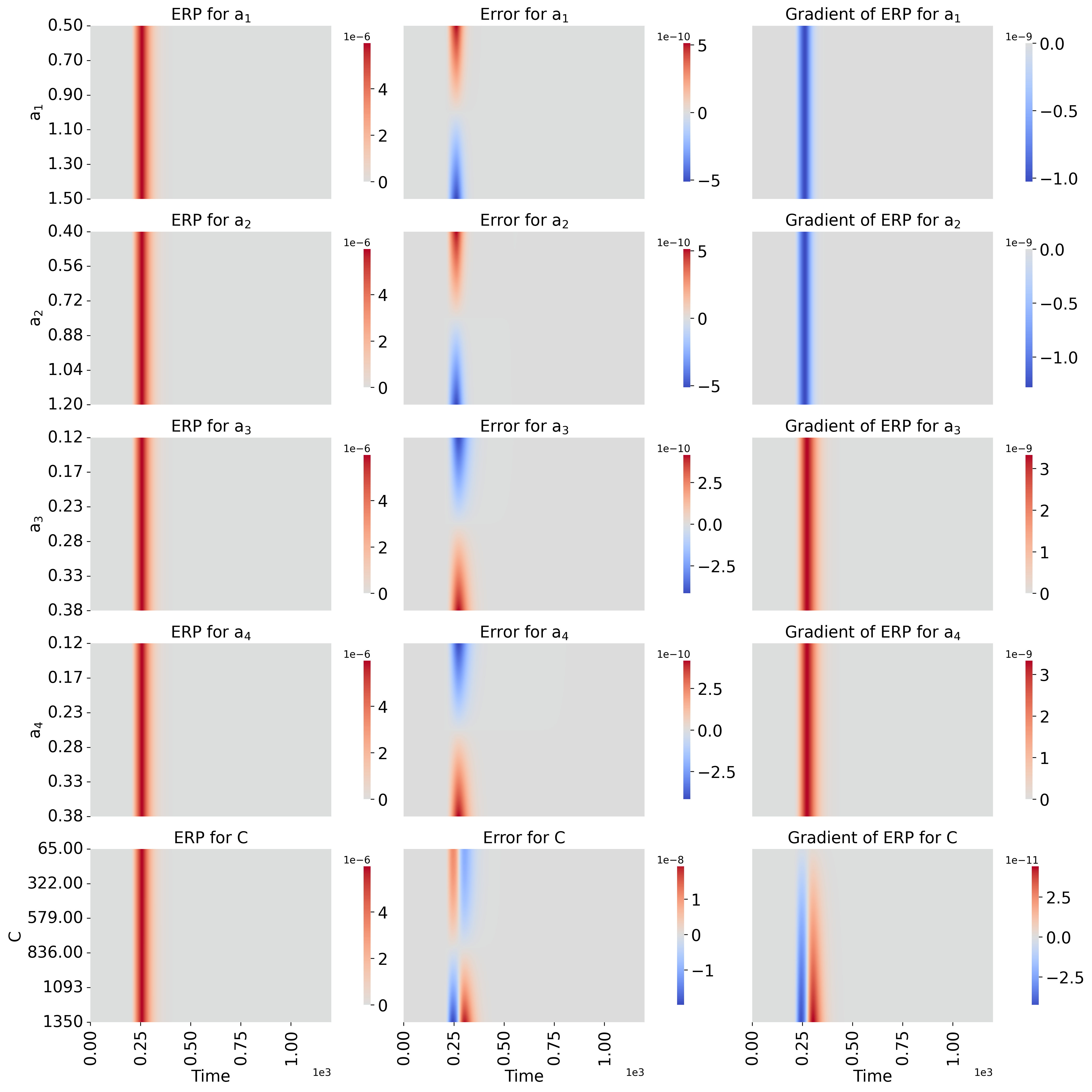}
    \caption{Sensitivity analysis for local connectivity ($a_1$-$a_4$) and C showing the ERP amplitude (left), error (middle), and gradient (right). 
    }
    \label{fig:sensitivity_group2}
    \vspace{-6mm}
\end{figure}

Sensitivity analysis revealed marked differences in parameter influence on simulated ERPs ( \autoref{fig:sensitivity_group1} and \autoref{fig:sensitivity_group2}). These sensitivity heatmaps highlight how each parameter shapes the temporal profile of the ERP. The raw ERP demonstrated that excitatory parameters ($A_e$, $a_1$, $a_2$) primarily increased the peak amplitude around 250ms, while inhibitory parameters ($A_i$, $a_3$, $a_4$) have an opposite effect. The error (deviation from the mean response) exhibited clear parameter-specific temporal windows of influence: $A_e$ and $A_i$ showed unimodal patterns whereas $b_e$ and $b_i$ showed an alternance of positive and negative peaks. The gradient further quantified sensitivity magnitude, revealing that $A_i$ had a more significant influence (about an order of magnitude) than $A_e$ on ERP morphology, despite both being gain parameters. Key takeaways include: (1) connectivity parameters exhibit functional differentiation, with \{$a_1$, $a_2$\} and \{$a_3$, $a_4$\} showing opposite gradient polarities; (2) the global coupling parameter $C$ affects multiple aspects of ERP morphology with complex biphasic patterns; and (3) excitatory time constant $b_e$ demonstrated distributed temporal effects (i.e., longer tail), 
indicating its crucial role in overall signal timing. These findings provide essential insights for parameter estimation and model interpretation in neurophysiological studies.

\vspace{-1mm}

\subsection{Parameter Inference Performance}
\begin{figure}[!ht]
   \centering
   \includegraphics[width=\columnwidth]{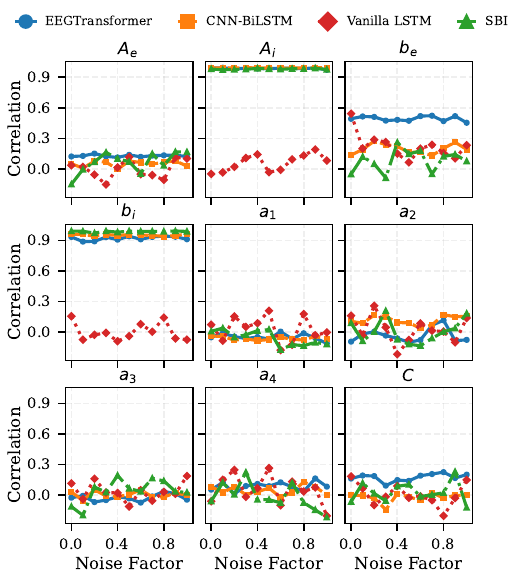}
   \vspace{-6mm}
   \caption{Comparison of deep learning models estimating JR-NMM parameters across noise levels. EEGTransformer and CNN-BiLSTM maintaining high correlations for $b_i$. The vanilla LSTM consistently underperforms with lower correlations, while SBI exhibits parameter-dependent variability.}
   \label{fig:correlation_results}
   \vspace{-6mm}
\end{figure}

The parameter inference capabilities of our four models—EEGTransformer, CNN-BiLSTM, Vanilla LSTM, and SBI—were evaluated across varying noise conditions (\autoref{fig:correlation_results}). Results show distinct patterns of parameter recoverability across the different architectures. The inhibitory parameter $b_i$ demonstrated the strongest recovery performance across multiple models. The EEGTransformer and CNN-BiLSTM maintained exceptionally high correlation coefficients (>0.9) for this parameter even at high noise levels, with SBI showing similarly strong performance. In contrast, the Vanilla LSTM performed poorly on this parameter, with correlations fluctuating around 0.1-0.2, suggesting fundamental limitations in capturing temporal dependencies (i.e., adding a 1D CNN to meaningful temporal features allowed the CNN-BiLSTM to perform much better than the vanilla LSTM). The excitatory parameter $b_e$ showed moderate recoverability, with the EEGTransformer achieving consistent correlations of approximately 0.5 across all noise conditions. The CNN-BiLSTM and SBI showed more variable performance for this parameter, with correlations typically in the 0.1-0.3 range, while Vanilla LSTM exhibited inconsistent correlations between 0.1-0.5. The connectivity parameters ($a_1$-$a_4$) exhibited poor recoverability across all models, with correlations rarely exceeding 0.2 and often oscillating around zero. Parameter $C$ showed better recoverability with the EEGTransformer than other models, maintaining positive correlations around 0.2 across noise levels.

The EEGTransformer and CNN-BiLSTM demonstrated superior robustness to noise compared to both the Vanilla LSTM and SBI approaches.

\section{Discussion}

Our comprehensive benchmarking study offers several key insights into the challenging problem of NMMs parameter inference from EEG data. NMM parameter estimation allows us to study the underlying neurophysiological mechanisms generating EEG. By estimating JR-NMM parameters from EEG, we directly map measurable brain signals and the biophysical properties of neural populations, enabling a mechanistic understanding of cortical dynamics and the potential identification of aberrant parameter configurations in pathological conditions. The sensitivity analysis revealed a clear hierarchy of parameter identifiability within the JR-NMM, with synaptic gains and time constants exhibiting stronger ERP signatures than connectivity parameters. This finding aligns with the fundamental architecture of the JR-NMM, where synaptic parameters directly modulate signal amplification and decay, creating more distinctive output patterns. The sensitivity analysis also clarifies why some parameters show poor identifiably. For example, $A_i$ and $A_e$  have almost identical patterns, only with inverse amplitude. Therefore, it is difficult to determine whether a reduction in ERP amplitude is caused by a decrease in $A_e$ or an increase in $A_i$. Consequently, in terms of identifiability, the JR model may provide better results for inverse modeling if we reparameterize it as a function of the excitatory-inhibitory ratio $r_{ei} = A_e/A_i$ and a global offset parameter $\alpha = (A_e+A_i)/2$ such that $A_e = 2\alpha r_{ei}/(1+r_{ei})$ and $A_i = 2\alpha/(1 + r_{ei})$. Moreover, such a parameterization would map well with the report of an imbalance in the excitatory/inhibitory ratio in various conditions, including autism spectrum disorder \cite{uzunova2016excitatory}.
\\
The comparative analysis of inference approaches demonstrated that Transformer-based architectures achieve superior robustness to noise compared to both the LSTM and SBI approaches. This advantage likely stems from the Transformer's self-attention mechanism, which can capture non-sequential dependencies across the entire ERP signal, unlike the strictly sequential processing of LSTMs. The consistent performance of the EEGTransformer, particularly for $A_i$ and $b_i$ parameters, suggests that attention-based architectures may be better suited for extracting relevant features from complex temporal signals like EEG.

The parameter recovery performance broadly corresponds with our sensitivity analysis findings, with parameters showing minimal impact on ERP morphology ($a_1$-$a_4$) being poorly recovered by all models, regardless of noise level. This observation confirms the fundamental challenge of parameter identifiability in complex dynamical systems. If parameters do not create distinct signatures in the observable output, no inference method can reliably recover them without additional constraints or information. The SBI approach performed well for some parameters but is highly sensitive to noise for others, suggesting that SBI may be valuable in low-noise conditions or combined with other ensemble-method approaches. The LSTM model's inconsistent performance across noise levels indicates potential limitations in applying recurrent architectures to this problem without additional regularization or architectural modifications.

We must acknowledge several limitations to this study. First, it used simulated data that incorporated realistic noise but could not capture the full complexity of real EEG recordings. Furthermore, we simulated neural activity using a JR-NMM in a single cortical region, while brain activity typically involves distributed networks of interacting regions. The parameter identifiability patterns observed may differ in multi-region models or when using alternative neural mass formulations with different state variables and connectivity architectures.

\section{Conclusion}

This study demonstrates that Transformer-based and CNN-BiLSTM architectures outperform LSTM and SBI approaches for JR-NMM parameter inference, with synaptic parameters showing higher recoverability than connectivity parameters. Our sensitivity analysis establishes a clear relationship between parameter influence on ERP morphology and inference reliability, providing a foundation for more robust approaches for NMMs parameter estimation. These insights have important implications for computational neuroscience and clinical applications. Future work should extend this analysis to empirical EEG data, explore transfer learning approaches to bridge synthetic and real data domains, and investigate the potential of hybrid architectures that combine the strengths of different inference approaches. Expanding this benchmarking to other neural mass models and exploring multi-modal data integration could enhance parameter identifiability. Further, reparameterizing existing NMM may offer equivalent but more identifiable models. Robust parameter inference methods will ultimately enable more reliable neural dynamics characterization in healthy and pathological states, advancing our understanding of brain function and dysfunction.


\bibliography{references}

\begin{thebibliography}{27}
\providecommand{\natexlab}[1]{#1}
\providecommand{\url}[1]{\texttt{#1}}
\expandafter\ifx\csname urlstyle\endcsname\relax
  \providecommand{\doi}[1]{doi: #1}\else
  \providecommand{\doi}{doi: \begingroup \urlstyle{rm}\Url}\fi

\bibitem[Anwar et~al.(2024)Anwar, Khalifa, Coyle, and Sejdic]{anwar2024transformers}
A.~Anwar, Y.~Khalifa, J.L. Coyle, and E.~Sejdic.
\newblock Transformers in biosignal analysis: A review.
\newblock \emph{Information Fusion}, page 102697, 2024.

\bibitem[Cakan et~al.(2021)Cakan, Jajcay, and Obermayer]{cakan2021}
C.~Cakan, N.~Jajcay, and K.~Obermayer.
\newblock neurolib: A simulation framework for whole-brain neural mass modeling.
\newblock \emph{Cognitive Computation}, Oct 2021.
\newblock ISSN 1866-9964.

\bibitem[Cranmer et~al.(2020)Cranmer, Brehmer, and Louppe]{cranmer2020frontier}
K.~Cranmer, J.~Brehmer, and G.~Louppe.
\newblock The frontier of simulation-based inference.
\newblock \emph{Proceedings of the National Academy of Sciences}, 117\penalty0 (48):\penalty0 30055--30062, 2020.

\bibitem[David and Friston(2003)]{articleDavidNMM}
O.~David and K.~Friston.
\newblock A neural mass model for {MEG/EEG}: coupling and neuronal dynamics.
\newblock \emph{NeuroImage}, 20:\penalty0 1743--55, 12 2003.

\bibitem[Deco et~al.(2008)Deco, Jirsa, Robinson, Breakspear, and Friston]{deco2008dynamic}
G.~Deco, V.K. Jirsa, P.A. Robinson, M.~Breakspear, and K.~Friston.
\newblock The dynamic brain: from spiking neurons to neural masses and cortical fields.
\newblock \emph{PLoS computational biology}, 4\penalty0 (8):\penalty0 e1000092, 2008.

\bibitem[Gelman and Hill(2006)]{gelman2006data}
A.~Gelman and J.~Hill.
\newblock \emph{Data analysis using regression and multilevel/hierarchical models}.
\newblock Cambridge University Press, 2006.

\bibitem[Greenberg et~al.(2019)Greenberg, Nonnenmacher, and Macke]{greenberg2019automatic}
D.~Greenberg, M.~Nonnenmacher, and J.~Macke.
\newblock Automatic posterior transformation for likelihood-free inference.
\newblock In \emph{International Conference on Machine Learning}, pages 2404--2414. PMLR, 2019.

\bibitem[Griffiths et~al.(2022)Griffiths, Wang, Ather, Momi, Rich, Diaconescu, McIntosh, and Shen]{GriffithsDCMfmri}
J.D. Griffiths, Z.~Wang, S.H. Ather, D.~Momi, S.~Rich, A.~Diaconescu, A.R. McIntosh, and K.~Shen.
\newblock Deep learning-based parameter estimation for neurophysiological models of neuroimaging data.
\newblock \emph{bioRxiv}, pages 2022--05, 2022.

\bibitem[Herzog et~al.(2024)Herzog, Mediano, Rosas, Luppi, Sanz-Perl, Tagliazucchi, Kringelbach, Cofr{\'e}, and Deco]{herzog2024neural}
R.~Herzog, P.A.M. Mediano, F.E. Rosas, A.I. Luppi, Y.~Sanz-Perl, E.~Tagliazucchi, M.~L. Kringelbach, R.~Cofr{\'e}, and G.~Deco.
\newblock Neural mass modeling for the masses: Democratizing access to whole-brain biophysical modeling with fastdmf.
\newblock \emph{Network Neuroscience}, 8\penalty0 (4):\penalty0 1590--1612, 2024.

\bibitem[Jansen and Rit(1995)]{JansenRit1995}
B.H. Jansen and V.G. Rit.
\newblock Electroencephalogram and visual evoked potential generation in a mathematical model of coupled cortical columns.
\newblock \emph{Biological Cybernetics}, 1995.

\bibitem[Kuhlmann et~al.(2016)Kuhlmann, Freestone, Manton, Heyse, Vereecke, Lipping, Struys, and Liley]{kuhlmann2016neural}
L.~Kuhlmann, D.R. Freestone, J.H. Manton, B.~Heyse, H.E.M. Vereecke, T.~Lipping, M.M.R.F. Struys, and D.T.J. Liley.
\newblock Neural mass model-based tracking of anesthetic brain states.
\newblock \emph{NeuroImage}, 133:\penalty0 438--456, 2016.

\bibitem[Lopes~da Silva et~al.(1974)Lopes~da Silva, Hoeks, Smits, and Zetterberg]{lopes1974model}
F.H. Lopes~da Silva, A.~Hoeks, H.~Smits, and L.H. Zetterberg.
\newblock Model of brain rhythmic activity: the alpha-rhythm of the thalamus.
\newblock \emph{Kybernetik}, 15:\penalty0 27--37, 1974.

\bibitem[Moran et~al.(2013)Moran, Pinotsis, and Friston]{moran2013neural}
Rosalyn Moran, Dimitris~A Pinotsis, and Karl Friston.
\newblock Neural masses and fields in dynamic causal modeling.
\newblock \emph{Frontiers in computational neuroscience}, 7:\penalty0 57, 2013.

\bibitem[Nunez and Srinivasan(2006)]{nunez2006electric}
P.L. Nunez and R.~Srinivasan.
\newblock \emph{Electric fields of the brain: the neurophysics of EEG}.
\newblock Oxford University Press, 2006.

\bibitem[Papamakarios and Murray(2016)]{papamakarios2016fast}
G.~Papamakarios and I.~Murray.
\newblock Fast $\varepsilon$-free inference of simulation models with bayesian conditional density estimation.
\newblock \emph{Advances in neural information processing systems}, 29, 2016.

\bibitem[Pinotsis et~al.(2014)Pinotsis, Robinson, Beim~G., and Friston]{articleNMF}
D.~Pinotsis, P.~Robinson, Peter Beim~G., and K.~Friston.
\newblock Neural masses and fields: Modelling the dynamics of brain activity.
\newblock \emph{Frontiers in Computational Neuroscience}, 8, 10 2014.

\bibitem[Sadeghi et~al.(2020)Sadeghi, Mier, Gerchen, Schmidt, and Hass]{sadeghi2020dynamic}
S.~Sadeghi, D.~Mier, M.F. Gerchen, S.N.L. Schmidt, and J.~Hass.
\newblock {Dynamic Causal Modeling for fMRI} with {Wilson-Cowan-based} neuronal equations.
\newblock \emph{Frontiers in Neuroscience}, 14:\penalty0 593867, 2020.

\bibitem[Sanz~Leon et~al.(2013)Sanz~Leon, Knock, Woodman, Domide, Mersmann, McIntosh, and Jirsa]{sanz2013virtual}
P.~Sanz~Leon, S.A. Knock, M.M. Woodman, L.~Domide, J.~Mersmann, A.R. McIntosh, and V.~Jirsa.
\newblock The virtual brain: a simulator of primate brain network dynamics.
\newblock \emph{Frontiers in neuroinformatics}, 7:\penalty0 10, 2013.

\bibitem[Sotero et~al.(2007)Sotero, Trujillo-Barreto, Iturria-Medina, Carbonell, and Jimenez]{sotero2007realistically}
R.C. Sotero, N.J. Trujillo-Barreto, Y.~Iturria-Medina, F.~Carbonell, and J.C. Jimenez.
\newblock Realistically coupled neural mass models can generate eeg rhythms.
\newblock \emph{Neural computation}, 19\penalty0 (2):\penalty0 478--512, 2007.

\bibitem[Spiegler et~al.(2011)Spiegler, Kn{\"o}sche, Schwab, Haueisen, and Atay]{spiegler2011modeling}
A.~Spiegler, T.R. Kn{\"o}sche, K.~Schwab, J.~Haueisen, and F.M. Atay.
\newblock Modeling brain resonance phenomena using a neural mass model.
\newblock \emph{PLoS computational biology}, 7\penalty0 (12):\penalty0 e1002298, 2011.

\bibitem[Stefanovski et~al.(2021)Stefanovski, Meier, Pai, Triebkorn, Lett, Martin, B{\"u}lau, Hofmann-Apitius, Solodkin, McIntosh, et~al.]{stefanovski2021bridging}
L.~Stefanovski, J.M. Meier, R.K. Pai, P.~Triebkorn, T.~Lett, L.~Martin, K.~B{\"u}lau, M.~Hofmann-Apitius, A.~Solodkin, A.R. McIntosh, et~al.
\newblock Bridging scales in alzheimer's disease: Biological framework for brain simulation with the virtual brain.
\newblock \emph{Frontiers in Neuroinformatics}, 15:\penalty0 630172, 2021.

\bibitem[Tejero-Cantero et~al.(2020)Tejero-Cantero, Boelts, Deistler, Lueckmann, Durkan, Gonçalves, Greenberg, and Macke]{tejero-cantero2020sbi}
A.~Tejero-Cantero, J.~Boelts, M.~Deistler, J.-M. Lueckmann, C.~Durkan, P.J. Gonçalves, D.S. Greenberg, and J.H. Macke.
\newblock sbi: A toolkit for simulation-based inference.
\newblock \emph{Journal of Open Source Software}, 5\penalty0 (52):\penalty0 2505, 2020.

\bibitem[Uzunova et~al.(2016)Uzunova, Pallanti, and Hollander]{uzunova2016excitatory}
G.~Uzunova, S.~Pallanti, and E.~Hollander.
\newblock Excitatory/inhibitory imbalance in autism spectrum disorders: Implications for interventions and therapeutics.
\newblock \emph{The World Journal of Biological Psychiatry}, 17\penalty0 (3):\penalty0 174--186, 2016.

\bibitem[Wang et~al.(2023)Wang, Leja, Villar, and Speagle]{wang2023sbi++}
B.~Wang, J.~Leja, V.A Villar, and J.S. Speagle.
\newblock Sbi++: Flexible, ultra-fast likelihood-free inference customized for astronomical applications.
\newblock \emph{The Astrophysical Journal Letters}, 952\penalty0 (1):\penalty0 L10, 2023.

\bibitem[Wendling et~al.(2016)Wendling, Benquet, Bartolomei, and Jirsa]{wendling2016computational}
F.~Wendling, P.~Benquet, F.~Bartolomei, and V.~Jirsa.
\newblock Computational models of epileptiform activity.
\newblock \emph{Journal of neuroscience methods}, 260:\penalty0 233--251, 2016.

\bibitem[Wilson and Cowan(1972)]{Wilson1972ExcitatoryAI}
H.R. Wilson and J.D. Cowan.
\newblock Excitatory and inhibitory interactions in localized populations of model neurons.
\newblock \emph{Biophysical journal}, 12 1:\penalty0 1--24, 1972.

\bibitem[Yu et~al.(2019)Yu, Si, Hu, and Zhang]{yu2019review}
Y.~Yu, X.~Si, C.~Hu, and J.~Zhang.
\newblock A review of recurrent neural networks: Lstm cells and network architectures.
\newblock \emph{Neural computation}, 31\penalty0 (7):\penalty0 1235--1270, 2019.

\end{thebibliography}

\end{document}